\documentclass{Nature}
\usepackage[normalem]{ulem}
\usepackage{graphicx}
\makeatletter
\let\saved@includegraphics\includegraphics
\AtBeginDocument{\let\includegraphics\saved@includegraphics}
\renewenvironment*{figure}{\@float{figure}}{\end@float}
\usepackage{color}
\usepackage{amsmath}
\makeatletter

\makeatletter

\newcommand{\diff}{{\mathrm d}}

\newcommand{\maya}{\textsc{Maya}}
\newcommand{\etk}{\textsc{EinsteinToolkit}}
\newcommand{\carpet}{\textsc{Carpet}}
\newcommand{\kranc}{\textsc{Kranc}}
\newcommand{\cactus}{\textsc{Cactus}}
\newcommand{\ahf}{\textsc{AHFinderDirect}}

\def\gw#1{gravitational wave#1 (GW#1)\gdef\gw{GW}}
\def\bh#1{black hole#1 (BH#1)\gdef\bh{BH}}
\def\bbh#1{binary black hole#1  (BBH#1)\gdef\bbh{BBH}}

\def\ligo#1{Laser Interferometer Gravitational-wave Observatory#1 (LIGO#1)\gdef\ligo{LIGO}}
\def\lisa#1{the Laser Interferometer Space Antenna#1 (LISA#1)\gdef\lisa{LISA}}
\def\et#1{the Einstein Telescope#1 (ET#1)\gdef\et{ET}}

\makeatother

\title{Post-merger chirps from binary black holes as probes of the final black-hole horizon}

\author{Juan Calderon Bustillo$^{\dagger, 1,2,3,4}$, Christopher Evans$^{\ddagger, 4}$, James A. Clark$^4$, Grace Kim$^{4,5}$, Pablo Laguna$^{4,6}$ and Deirdre Shoemaker$^{4,6}$}

\begin{document}

\maketitle

\begin{affiliations}
\item Department of Physics, The Chinese University of Hong Kong, Shatin, N.T., Hong Kong
\item Monash Centre for Astrophysics, School of Physics and Astronomy, Monash University, VIC 3800, Australia
\item OzGrav: The ARC Centre of Excellence for Gravitational-Wave Discovery, Clayton, VIC 3800, Australia
\item Center for Relativistic Astrophysics and School of Physics, Georgia Institute of Technology, Atlanta, GA 30332
\item Department of Physics and Astronomy, Stony Brook University, Stony Brook NY 11794, USA
\item Center for Gravitational Physics, Department of Physics, University of Texas at Austin, Austin, TX 78712, USA
\end{affiliations}

\section{Abstract}
\begin{abstract}
The merger of a binary black hole gives birth to a highly distorted final black hole. The gravitational radiation emitted as this black hole relaxes presents us with the unique opportunity to probe extreme gravity and its connection with the dynamics of the black hole horizon. Using numerical relativity simulations, we demonstrate a connection between a concrete observable feature in the gravitational waves and geometrical features on the dynamical apparent horizon of the final black hole. Specifically, we show how the line-of-sight passage of a ``cusp''-like defect on the  horizon of the final black hole correlates with ``chirp''-like frequency peaks in the post-merger gravitational-waves. These post-merger chirps should be observed and analyzed as the sensitivity of LIGO and Virgo increases and as future generation detectors, such as LISA and the Einstein Telescope, become operational. 
\end{abstract}

\section{Introduction}

A new field of astronomy has arisen with the detection of \gw{s}. To date, the \ligo{} \cite{TheLIGOScientific:2014jea} and Virgo \cite{TheVirgo:2014hva} have observed twelve merging \bbh{s} ~\cite{Abbott:2016blz,GWTC1,LIGOScientific:2020stg,GW190412,Abbott:2020uma,GW190521D}, two binary neutron star mergers ~\cite{TheLIGOScientific:2017qsa,Abbott_2020} and a putative neutron star-black hole merger \cite{NSBH} . These detections are 
allowing us to understand the nature of compact objects, their populations~\cite{O1-O2-rate_paper} and their formation channels~\cite{TheLIGOScientific:2016htt,Stevenson:2017tfq}. These observations have also put to test General Relativity (GR)~\cite{Einstein:1916vd,Misner:1974qy,LIGOScientific:2019fpa}  in the strong field regime for the first time, so far confirming its predictions ~\cite{TheLIGOScientific:2016src,PhysRevX.6.041015}. Despite this groundbreaking achievement, \ligo{} and Virgo have not yet reached the sensitivity to observe in exquisite detail the merger and the relaxation of the highly distorted black holes (BHs) left behind by BBHs, 
when dynamical gravity reaches its ultimate expression. As the sensitivity of \ligo{} and Virgo increases and future generation detectors, such as \lisa{} and the \et{}, become operational, \gw{s} will provide us with an unprecedented view of highly distorted BH horizons, allowing us to test in further detail fundamental aspects of GR like the ``no-hair'' theorem~\cite{Misner:1974qy,Isi:2019aib} and explore the quantum properties of BHs~\cite{Cardoso:2016oxy}.

Such studies will rely on a deep understanding of how GW signals encode the dynamical properties of the source.
In anticipation of future large signal-to-noise ratio detections, it is important to investigate how \gw{s} reflect not only the ``common'' properties of BBHs (e.g., BH masses and spins, orbital eccentricity and orientation) but also other fundamental aspects that can be inferred from the morphology of the signal. For instance, all current observations show a rather simple ``chirp'' morphology \cite{GWTC1}, consisting of a monotonic increase of both frequency and amplitude.  
Initially, both quantities increase slowly, reflecting the low frequency and tightening of the orbit as the two BHs approach each other \cite{Maggiore:2007shp}. Just before merger, the two BHs reach speeds comparable to that of light, leading to a rapid rise of both frequency and amplitude~\cite{Maggiore:2007shp, Pretorius:2005gq}. Once the BHs merge, a highly distorted  BH settles into a Kerr BH radiating exponentially decaying ringdown emission ~\cite{Berti:2009kk,Owen:2009sb,Abbott:2016blz,Isi:2019aib}. However, this simple morphology describes the GW signal only when the binary has comparable masses or it is viewed face-on, so that the emission is vastly dominated by the so-called quadrupole mode.

In contrast, asymmetric 
BBHs also show strong \gw{} emission in sub-dominant higher order modes during the merger and ringdown stages ~\cite{Berti:2007fi}, allowing for \gw{s} with non-trivial morphology complexity~\cite{Gonzalez:2008bi,Graff:2015bba,CalderonBustillo:2018zuq} that may unveil new features of the post-merger dynamics. The connection between the horizon dynamics and the \gw{} emission has been widely studied in two main ways. The first consists of finding correlations between far-field signals and fields near the horizon, revealing close connections between the horizon geometry, the \gw{} flux, and strong-field phenomena such as the anti-kick~\cite{Rezzolla:2010df,Jaramillo:2011re,Jaramillo:2011rf,Jaramillo:2012rr}. The second approach has focused on the systematic development of analytical tools that can probe and explain the geometrodynamics of spacetime causing such correlations and relate it with the generation of GWs \cite{CaltechPRL,Caltech1,Caltech2,Caltech3,Prasad:2020}. None of these studies discuss a direct link to the \gw{} strain observable by detectors. In this work, we follow the approach of \cite{Rezzolla:2010df,Jaramillo:2011re,Jaramillo:2011rf,Jaramillo:2012rr} to correlate a concrete observable feature of the GW strain to a geometrical property of the final BH horizon.\\

Here we present an explicit example of a complex post-merger signal and how it correlates with the dynamics of the evolving final black-hole horizon. Using numerical relativity simulations, we show that multiple post-merger frequency peaks (or chirps) can be measured near the orbital plane of unequal-mass binaries
We show that these correlate to the line-of-sight passage of strongly emitting regions of large mean curvature gradient and locally extremal Gaussian curvature, present on the dynamical apparent horizon of the final \bh{}, which cluster around a ``cusp''-like defect on it. 
Conversely, frequency minima correlate to the passage of the ``smoother'' opposite region of the horizon, where curvature gradients are smaller. \\   



\begin{center}
\begin{figure*}
\includegraphics[width=1\textwidth]{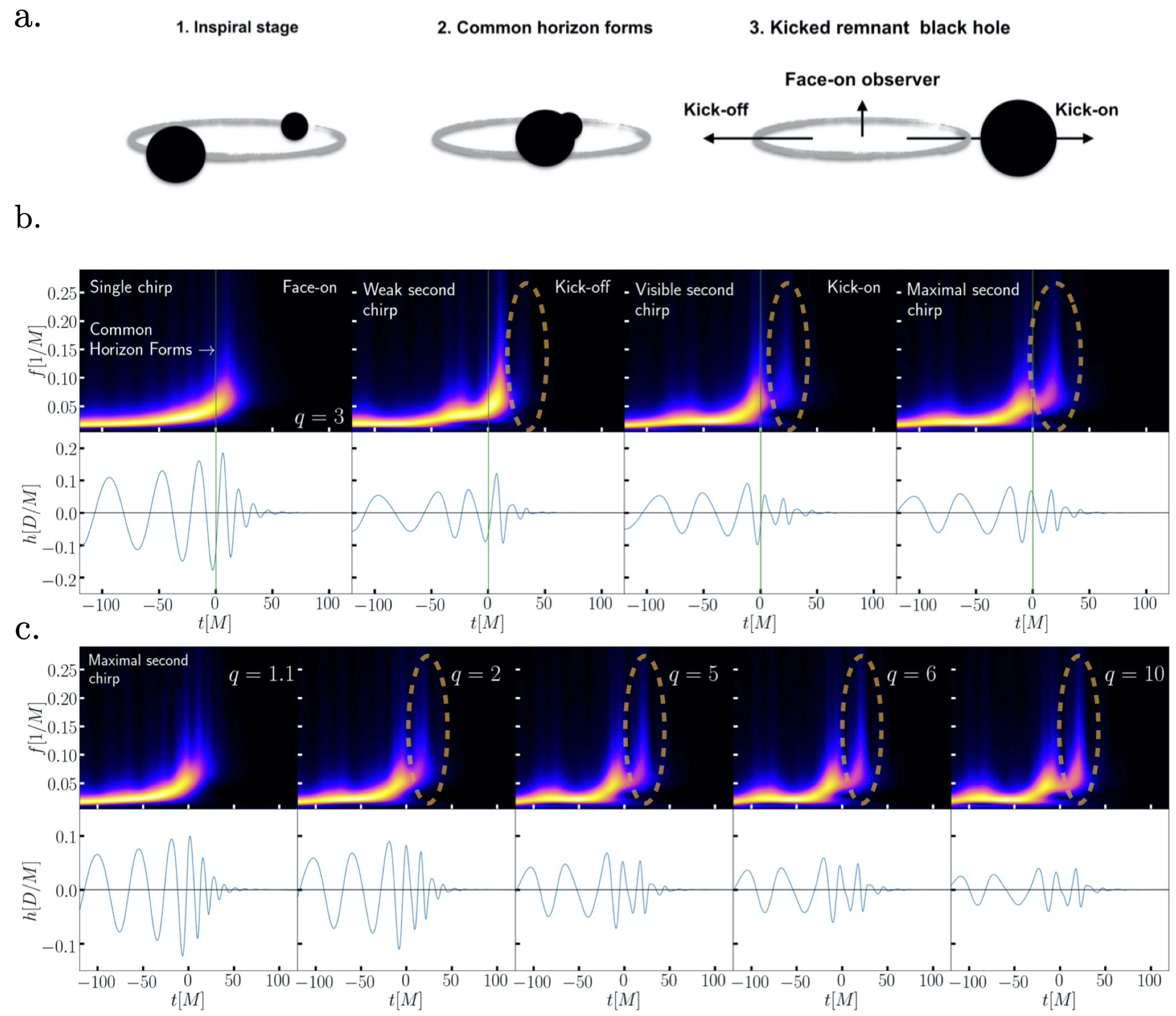}
\caption{\textbf{Post-merger chirps}. Panel a depicts the different stages of a binary black hole coalescence. The panels b and c show the strain time-series $h(t)$ of extracted from binary black hole simulations (white background) and the corresponding time-frequency maps (black background). The green vertical line denotes the instant at which the final common horizon is first found in our simulations. All waveforms include the most dominant modes $(\ell,m)=\{(2,\pm1),(2,\pm2),(3,\pm2),(3,\pm3),(4,\pm3),(4,\pm4)\}$. The b panels show the case of mass ratio $q=3$ for different viewing angles (from left to right): face-on, kick-off, kick-on and $55^\circ$ away from the kick direction measured in the direction of the orbit, for which the double-chirp feature is most prominent. The c panels show the signals emitted by binaries with a corresponding mass-ratio $q=1.1$ to $q=10$ in this last direction. The double-chirp feature becomes more pronounced as the mass ratio $q$ increases.}
\label{fig:waveforms}
\end{figure*}
\end{center}


\begin{center}
\begin{figure*}
\includegraphics[width=1\textwidth]{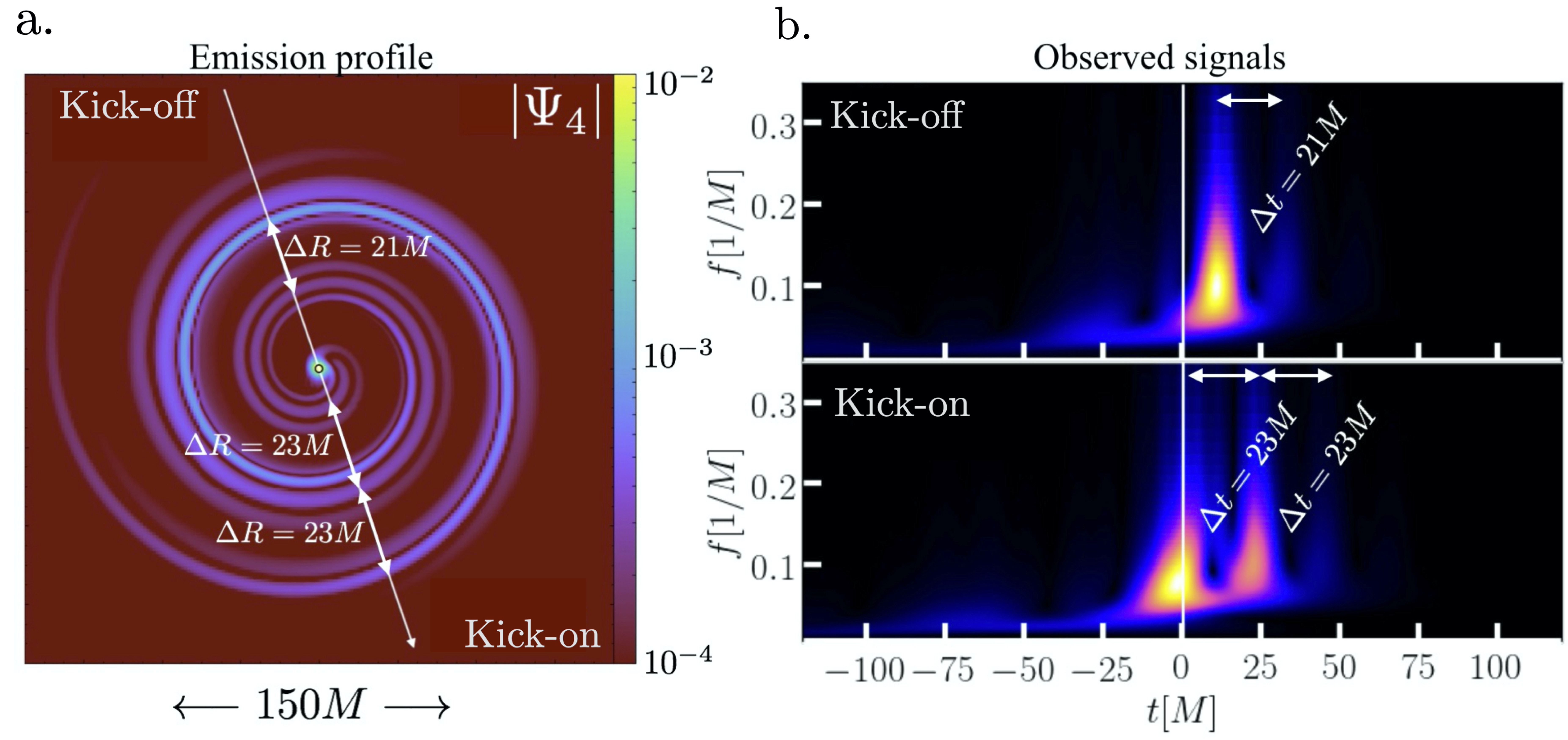}
\caption{\textbf{Relation between post-merger chirps and arriving gravitational-wave trains}.  Panel a shows a snapshot of the gravitational waves in the orbital plane at a time of $52.3M$ after the merger, for the  case of a $q=3$ binary. This is expressed in terms of the absolute value of the Newman-Penrose scalar $\Psi_4$. We highlight the kick-on and kick-off directions and the separation $\Delta t \sim \mathcal{O}(20M)$ of the wave-trains traveling towards each observer. Panel b shows the corresponding $\Psi_4$ time-frequency maps recorded by the kick-on and kick-off observers. The time elapsed  $\Delta t \sim \mathcal{O}(20M)$ between frequency peaks (or chirps) is consistent with the separation of the arriving wave-trains. Two strong wave-trains reach the kick-on observer, which translates into a prominent double-chirp. In contrast, the second front traveling towards the kick-off observer is much weaker, translating into a weak second chirp.}
\label{fig:far_zone}
\end{figure*}
\end{center}

\begin{center}
\begin{figure*}
\includegraphics[width=1\textwidth]{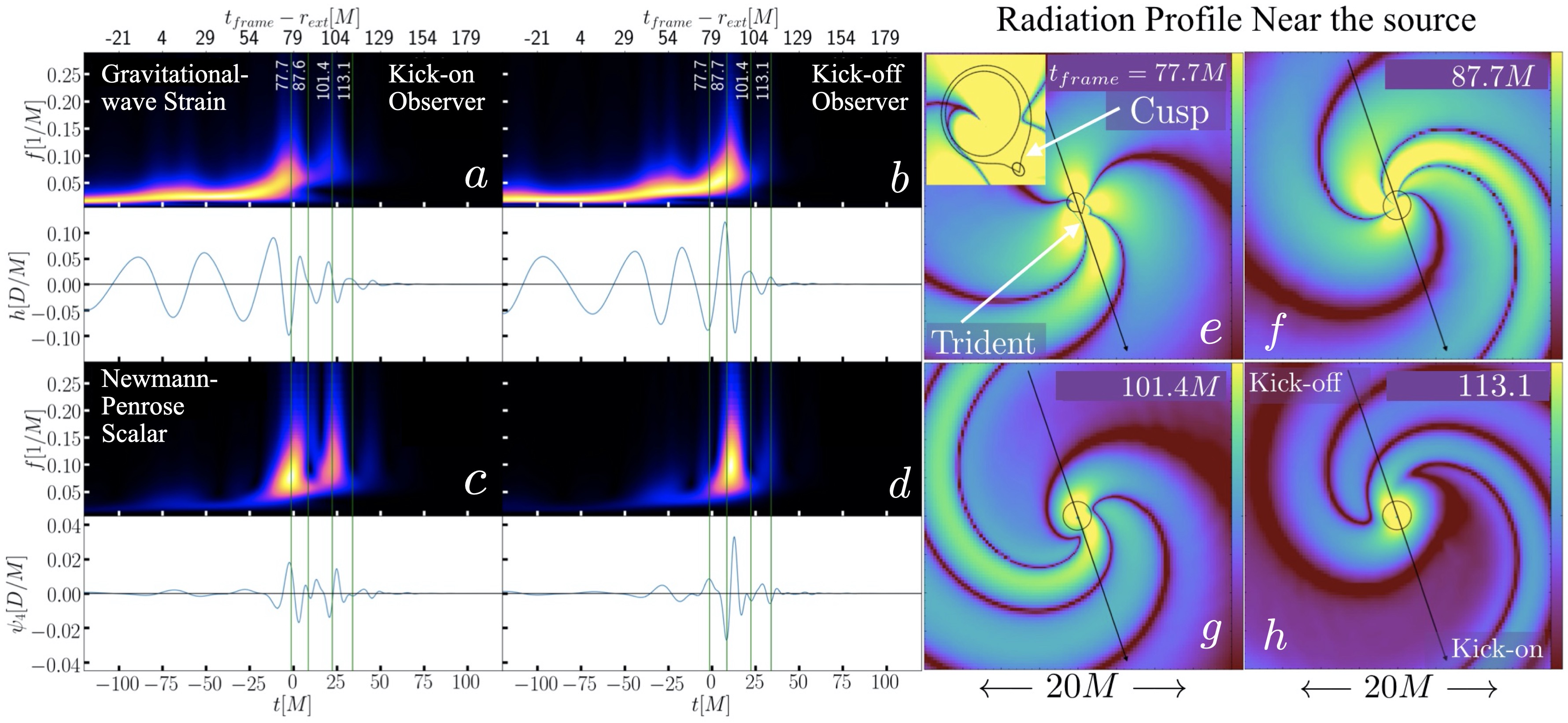}
\caption{\textbf{Relation between post-merger chirps and the black hole cusp passage through the line-of-sight}. Panels a and b show the time-frequency maps and time series of the gravitational-wave strain $h$, respectively observed in the kick-on and kick-off directions, extracted at a distance $r_{\text{ext}}=75M$ from the source. Panels c and d show the same for the Newman-Penrose scalar $\Psi_4$. The top time axis shows the retarded time $t_{\text{frame}}-r_{\text{ext}}$. The four vertical lines denote retarded times $t_{\text{frame}}-r_{\text{ext}} = t_{\text{frame},i}$, with the times  $t_{\text{frame},i}$ corresponding to the four simulation snapshots shown in panels e--h. The bottom time axis has been shifted so that $t=0$ denotes the time at which the $(2,2)$ emission mode has its amplitude peak, as it is common in GW data analysis. Panels e--h show the corresponding four frames  of the absolute value of $\Psi_4$ in the orbital plane of a $q=3$ binary, at the times $t_{\text{frame},i}$ highlighted in panels a--d. Bright yellow regions denote large values of $|\Psi_4|$ while dark purple regions denote zeros. The black arrow points in the kick-on direction. The inset highlights the initially asymmetric shape of the final black hole. The three-arm (or ``trident'') structure of $\Psi_4$ present on the bottom side of the horizon has its most prominent arm aligned with a ``cusp'' defect on it. }
\label{fig:retarded_times}
\end{figure*}
\end{center}

\section{Results and Discussion}

\subsection{Post-merger chirps}
\label{sec:waveform_morphology}
Figure~\ref{fig:waveforms}.a depicts the different stages of a \bbh{.} Fig. ~\ref{fig:waveforms}.b shows the GW strain time-series and time-frequency maps recorded by observers in different locations around a numerically simulated \bbh{} with a mass ratio $q = m_1/m_2 = 3$. We perform our numerical simulations using the \maya{} code, based on the \etk \cite{2012CQGLoffler} (for details, please see the Methods section). We use geometrical units in terms of the total mass of the binary $M=m_1+m_2$ with the speed of light set to unity. The green vertical line denotes the formation of the common apparent horizon. The ``face-on'' panel shows the signal observed face-on, showing the ``vanilla'' chirp structure consistent with current observations \cite{GWTC1,LIGOScientific:2020stg,GW190412}. The different panels in Fig.~\ref{fig:waveforms}.b  show the signals recorded at different positions on the orbital plane.\\ 
After the common horizon forms, the signal shows a clear drop in the frequency, followed by a secondary peak, or post-merger chirp. Depending on the location of the observer, the chirps occur at different times and involve different peak frequencies and intensities. It is illustrative to compare the signals observed in the direction of the recoil of the final BH \cite{Gonzalez:2008bi,CalderonBustillo:2018zuq}, (kick-on) to those observed in the opposite direction (kick-off). While the kick-on observer records a secondary chirp with larger amplitude and peak frequency than the first, the converse is measured by the kick-off observer. We find that the secondary chirp is more intense compared to the first one in a direction $\simeq 55^\circ$ away from the final recoil (or kick) direction, measured in the direction of the original orbit. This signal is shown in the rightmost panel, showing a clear ``double-chirp''. Notably, we find that this is also true for varying mass ratios. 
The corresponding signals are shown in Fig.~\ref{fig:waveforms}.c for binaries with mass ratios $q=1.1$ to $q=10$. These also 
make evident that the double-chirp signature becomes more pronounced as the binary becomes more asymmetric. We acknowledge that a similar non-trivial post-merger emission, visible in the time-domain, was shown by Gonz\'{a}lez et al., ~\cite{Gonzalez:2008bi} in terms of the Newman-Penrose scalar $\Psi_4$ . However, its frequency content, which is the departing point of our study and the reason behind the double-chirp name, was not shown.\\ 

Analytically, these complex and observer-dependent post-merger waveform morphologies can be explained by the asymmetric interaction of the different quasi-normal emission modes beyond the quadrupole one in different directions around the binary. These modes are triggered during the merger and ringdown of asymmetric binaries and have a larger impact for highly inclined binaries ~\cite{Berti:2007fi,CalderonBustillo:2018zuq,Giesler2019}. However, the clarity of the double-chirp signature suggests a connection to some underlying post-merger feature, similar to how the increase of the frequency during the inspiral is connected to the increasing frequency of the binary. In the following, we argue that this feature is the existence of regions of locally extremal curvature that are distributed non-uniformly on the dynamical apparent horizon of the final BH. While three of these regions cluster around a global curvature maximum that we denote as the ``cusp'' forming a ``trident'', a fourth one sits on the opposite or ``back'' side of the horizon. These coincide with regions of maximal GW emission. As the final BH relaxes, this structure rotates pointing to all observers on the orbital plane while fading away. We show that, after the cusp (back) of the horizon crosses the line-of-sight, frequency peaks (minima) are recorded at a time consistent with the GW travel time determined by the distance to the observer.\\

\subsection{Emission profile far from the source}

To make a first connection between the time-frequency morphology of the signal and the structure of the \gw{} emission, 
the left panel of Fig.~\ref{fig:far_zone}.a shows a snapshot of the GW emission in the orbital plane of our $q=3$ binary at a time of $52.3M$ after the merger, after the waves have traveled far from the source. We represent the GW emission  by the absolute value of the Newman-Penrose scalar $\Psi_4$\cite{Newman:1961qr}, related to the \gw{} strain $h$ by $\Psi_4(t) = \frac{d^2 h^{*}(t)}{dt^2}$. As one goes around the final \bh{}, it is clear that the recorded signal will depend on the viewing angle. This can be, in fact, observed in Fig.~\ref{fig:far_zone}.b, which showcases the time-frequency maps recorded by the kick-on and kick-off observers highlighted in the left panel. 
The radial separation of $\Delta R \sim \mathcal{O}(20M)$ between the wave-trains reaching each observer in the left panel is consistent 
with the time delay $\Delta t \sim \mathcal{O}(20M)$  between frequency peaks shown in the right one. 
Following the kick-on and kick-off directions, it is also evident that the intensity of the \gw{} fronts is different in each direction. While two wave-trains of similar intensity reach the kick-on observer, the second train reaching the kick-off observer has a much lower intensity. Consistently, the kick-on observer records two chirps of similar intensity plus a weak third one, while the second chirp is barely visible for the kick-off observer.

\begin{center}
\begin{figure}
\includegraphics[width=0.50\textwidth]{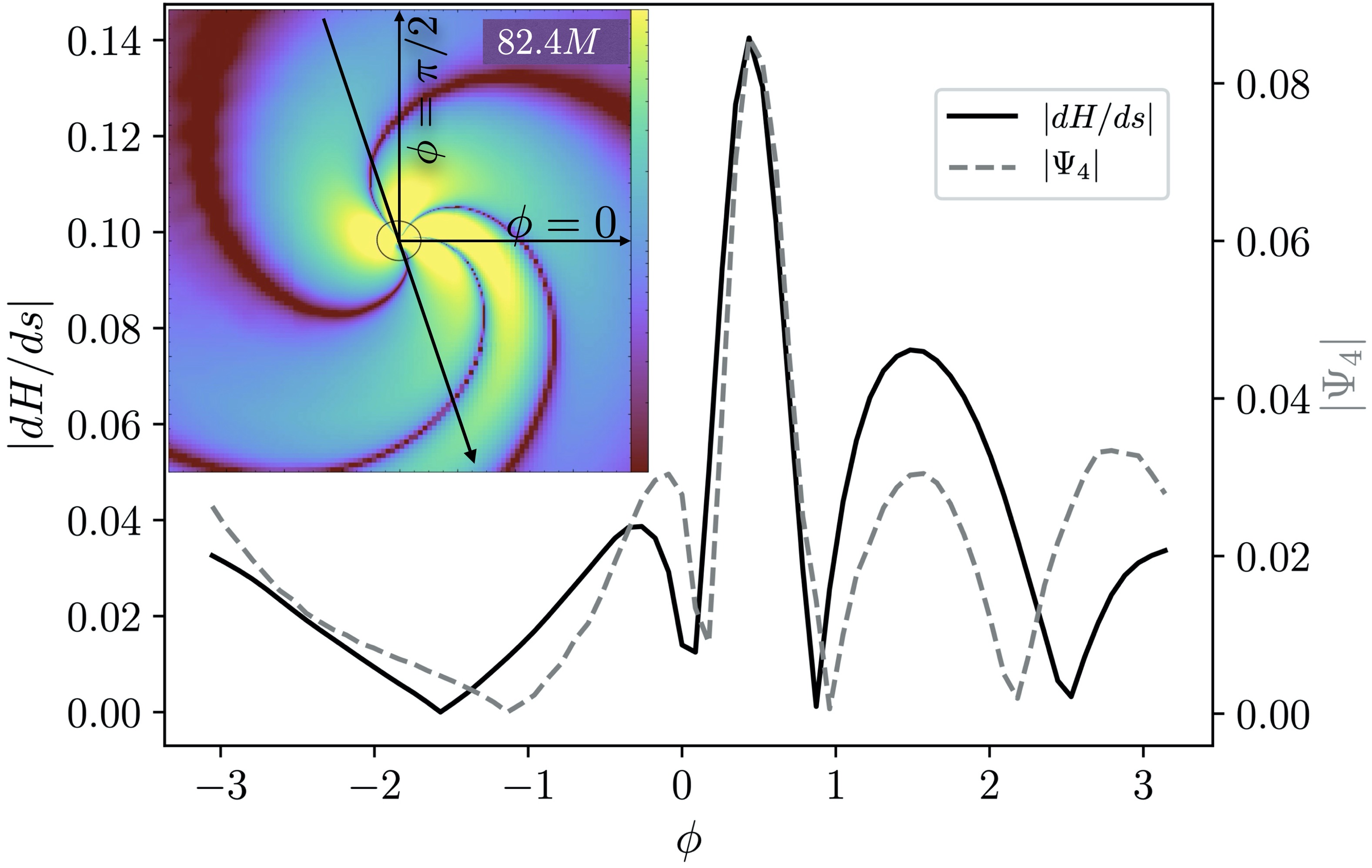}
\caption{\textbf{Relation between the gravitational-wave emission and mean curvature of the apparent horizon}. The main panel shows the absolute values of the gradient of the mean curvature $|dH/ds|$ and the Newman-Penrose scalar $|\Psi_4|$ measured on the intersection of the orbital plane and the final common apparent horizon of a $q=3$ binary as a function of the azimuthal angle $\phi$, measured $5M$ after the formation of of the common apparent horizon (corresponding to $t_{\text{frame}}=82.4$ in panels e--h of Fig. 3). The parameter $s$ denotes the arc-length along the equator. The inset shows the corresponding simulation snapshot for $|\Psi_4|$ measured on the orbital plane.}
\label{fig:curvature_and_psi}
\end{figure}
\end{center}

\begin{figure*}[t]
\begin{center}
\includegraphics[width=0.99\textwidth]{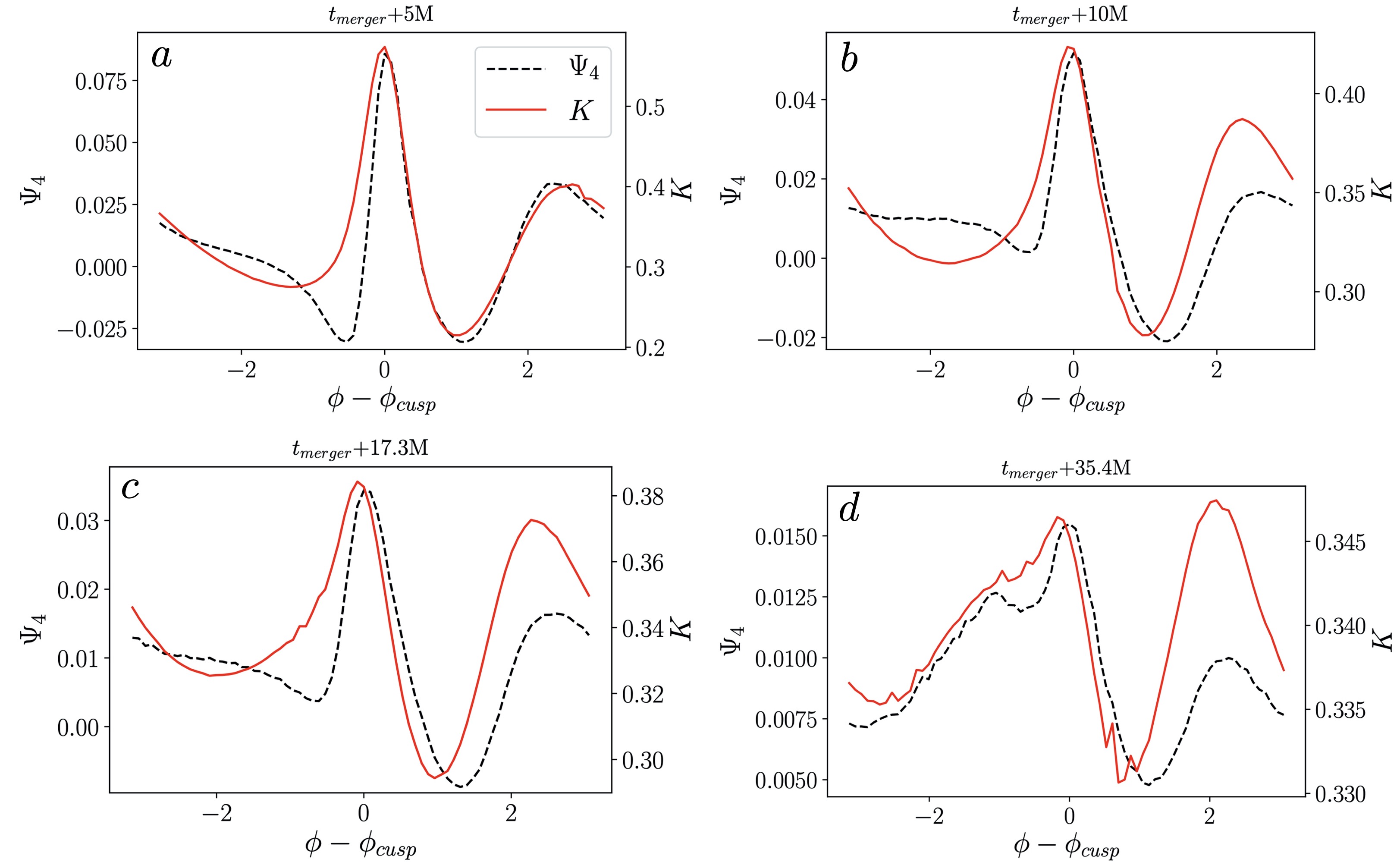}
\caption{\textbf{Relation between the gravitational-wave emission and Gaussian curvature of the apparent horizon}. Values of the Newman-Penrose scalar $\Psi_4$ and the Gaussian curvature $K$ on the intersection of the orbital plane and the final common apparent horizon of a $q=3$ binary as a function of the azimuthal angle $\phi$ at times $5M$, $10M$, $17.3M$ and $35.4M$ after the formation of the common apparent horizon. The $x$-axis has been shifted by a value $\phi_{\text{cusp}}$ so that the maximum of $\Psi_4$ (i.e., its central arm) is at $\phi_-\phi_{\text{cusp}}=0$.}
\label{fig:curvature_and_psi_time}
\end{center}
\end{figure*}

\subsection{The near horizon region}
After making a first connection between the post-merger chirps and the emission profile far from the source, we now zoom in near the source to investigate the connection to the post-merger dynamics. The panels e--h in Fig.\ref{fig:retarded_times} show the structure of $\Psi_4$ near the horizon at four selected times $t_{\text{frame}}$ throughout its evolution.  In the first frame, soon after the horizon forms, $\Psi_4$ shows a clearly asymmetric pattern. Three arms (a central one, most prominent, and two surrounding weaker ones) cluster on one side of the horizon forming a ``trident'' while another arm is present on the opposite side (or ``back''). We note that this structure does not form abruptly at merger but arises smoothly from the pattern it had when the \bh{}s where approaching each other. The inset shows that the central arm sits on a ``cusp'' defect present on the horizon. The other three frames show how this structure rotates and fades away as the final BH evolves. As this happens, the three arms (in particular the central one) and the back of the BH cross the line of sight of every observer multiple times.\\

We now draw our attention to the observed signals. The panels a--d in Fig. \ref{fig:retarded_times} show the time-frequency maps and time-series for the GW strain $h$ (panels a and b) and $\Psi_4$ (panels c and d) measured in the kick-on and kick-off directions at retarded times $t_{\text{frame}}-r_{\text{ext}}$, with $r_{\text{ext}}$ denoting the distance to the source, also known as the GW extraction radius. The vertical lines denote the retarded times that correspond to the four frames on Fig. \ref{fig:retarded_times}.b. As the three $\Psi_4$ arms, and in particular the central, most prominent one, cross the line-of-sight, a frequency peak is measured at a time $r_{\text{ext}}$ later. This is also noticeable in the time-domain plots in terms of a short instantaneous signal wavelength, consistent with the short separation of the arriving wave-fronts shown in Fig.2.a Similarly, a frequency minimum is observed $r_{\text{ext}}$ after the back of the horizon crosses the line of sight. This is reflected in the time-domain signals by a larger instantaneous wavelength, consistent with the large separation of ${\cal{O}}(20M)$ of the arriving wave-trains shown in Fig.2.a This way, we establish a time-connection between the line-of-sight passage of the three $\Psi_4$ arms (in particular the most prominent one) and the back arm, and the respective observation of post-merger frequency peaks and minima.\\

\subsection{Connecting the horizon geometry with post-merger chirps}

Next, we connect the arm structure of $\Psi_4$ present on the dynamical apparent horizon with its curvature. To this, we measure $\Psi_4$ on the intersection of the horizon with the orbital plane, together with the mean curvature $H$ and the Gaussian curvature $K$. Fig. \ref{fig:curvature_and_psi} shows the absolute values of the gradient of the mean curvature $dH/ds$ and $\Psi_4$ as a function of the azimuthal angle $\varphi$ at a time $5M$ after the common horizon has formed. The parameter $s$ denotes the arc-length along the horizon's equator (for a detailed description of the arc-length parameter and curvature quantities, please see Methods) and the inset panel shows the corresponding simulation frame. A tight correlation is obvious. In particular, all four $|\Psi_4|$ maxima, i.e., the arms highlighted in the inset, clearly correspond to local maxima of $|dH/ds|$. The three “tight” $\Psi_4$ arms correspond to three nearby regions of large $|dH/ds|$ spanning barely $\sim 2$ radians, with the maximum of $|dH/ds|$ (the cusp) matching the maximum of $|\Psi_4|$, i.e., the ``central'' arm. 

Fig.\ref{fig:curvature_and_psi_time}.a shows that the $\Psi_4$ arms also match regions of locally extremal Gaussian curvature $K$ which, unlike H, is intrinsic to the apparent horizon and coordinate independent. In Supplementary Fig. 1 and Supplementary Notes I, we show that the same relations hold for all the mass-ratios shown in Fig.\ref{fig:waveforms}. Moreover, while the location of the apparent horizon itself depends on the gauge choice, we show that these relations also hold in an alternative gauge. Panels b--d of Fig. \ref{fig:curvature_and_psi_time} and those in Supplementary Fig. 2 (and Supplementary Notes II), show that this relation is well preserved throughout the evolution of the final black hole, as the central arm points to the different observers in the orbital plane and the $\Psi_4$ structure fades away. We observe that a slight degradation of this correlation occurs at very late times of $\simeq 30M$ after the common horizon has formed, when the emission is very weak and more prone to be affected by numerical artefacts. Despite this, the three maxima and minima of $\Psi_4$ sitting around the central arm are well co-located with those of $K$. Also, while here we have used $|dH/ds|$ to facilitate a visual comparison with the simulation frames, we show that $\Psi_4$ actually follows $-dH/ds$.

Finally, even if not clearly visible Fig. 3 e--h,  Fig. \ref{fig:curvature_and_psi_time} makes also evident that the highly asymmetric structure of the horizon, understood as the ``clustering'' of three $\Psi_4$ arms and the corresponding extremal curvature regions, is maintained during BH evolution, as these always span an angle that barely exceeds $\sim 2.5$ radians. This way, the passage of these $\Psi_4$ arms by the line-of-sight  corresponds to the passage of three regions of extremal mean curvature gradient and extremal Gaussian curvature on the horizon, with the central arm corresponding to the cusp.

\subsection{Discussion}

The observation of the merger and ringdown stages of BH mergers grants access to the strongest regime of gravity, in which space-time shows its ultimate phenomenology through the dynamical evolution of highly distorted BH horizons. Connections between these dynamics and the observed GWs have been proposed and widely investigated ~\cite{Jaramillo:2011rf,Jaramillo:2011re,Jaramillo:2012rr,Rezzolla:2010df,CaltechPRL,Caltech1,Caltech2,Caltech3,Prasad:2020}. However, no explicit examples of concrete observable features in the GWs have been described to date.  In this work, we propose the first such connection.
We have shown that non-trivial features consisting on multiple frequency peaks in the post-merger GW emission of edge-on, asymmetric BH mergers, are linked to the presence of large curvature regions in the dynamical apparent horizon of the final BH that are asymmetrically distributed.\\ 
First, we have shown that an asymmetric emission pattern forms around the final common horizon, with three arms clustering one side and one on the other. Second, we have shown that frequency peaks, that we refer to as post-merger chirps, are observed in the post-merger signal as the three arms cross the line of sight, after a time consistent with the GW travel time. Last, we show that these arms coincide with locally extremal values of the Gaussian curvature and the mean curvature gradient of the dynamical apparent horizon, with the strongest arm sitting on its largest curvature region, which we call cusp. 
While post-merger chirps may resemble the signature of \bh{} echoes~\cite{Cardoso:2016oxy}, these are not, as we are considering standard BHs. For all the mass-ratios we have considered, we find this feature is more prominent on the orbital plane of the binary, $\simeq 55^\circ$ from the final kick direction measured in the direction of the original orbit (or that of the final spin). Nevertheless, we assume that this will be subject to change if spinning BHs were considered ~\cite{Apostolatos:1994,Kidder:1995}.\\
Finally, in the Supplementary Fig. 3 and Supplementary Notes III, we show that Advanced LIGO detectors working at their design sensitivity may observe the post-merger chirp signature for the case of a correctly oriented copy of the BBH GW170729 \cite{GWTC1,Chatziioannou2019}, suggesting that such observation may be feasible before the arrival of the next generation of GW detectors. 

\begin{methods}

\subsection{Time-frequency Maps:}\label{sec:cwt}
The continuous wavelet transform of a function $g(t)$ is,
\begin{equation}\label{eq:cwt}
   W(a,b) = \frac{1}{\sqrt{a}} \int_{-\infty}^{\infty} g(t)
   \psi\left(\frac{t-b}{a}\right)~\diff t,
\end{equation}
where $\psi$, the `mother wavelet`, is a continuous function in time and 
frequency, evaluated at scales $a>0$ and and translations $b$.  The scale 
parameter dilates the mother wavelet $\psi$, providing a range of 
time-frequency resolutions, while the translations provide time localization of 
the signal power, $|g^2(t)|$.  By selecting for $\psi$ a function which is 
compact in the time- and frequency-domains, together with a judicious 
choice of scales appropriate for the problem at hand, one can use the 
continuous wavelet transform to resolve substructure in the signal $g(t)$ of 
particular physical interest. A complete description of wavelet analysis may be 
found in~\cite{Walnut2004}.

The mother wavelet used in our decomposition is the Morlet
wavelet, a Gaussian-modulated sinusoid with minimal compactness in 
the time- and frequency-domains:
$
   \psi(x) = \frac{1}{\mathrm{\pi}^4} \exp \left( 2 \mathrm{\pi} i f_0 x\right) -
    \exp\left(-2\mathrm{\pi}^2 f_0^2 \right) \exp\left(x^2 / 2\right),
$
where $x=\frac{t-b}{a}$ and $f_0$ is the frequency of the sinusoid, the center 
frequency of the wavelet.

As described, the continuous wavelet transform is expressed in terms of mother 
wavelet dilation scales $a$.   Our interest lies in the frequency content of 
the signal, however.  It may be shown, by considering the response to a 
sinusoid of frequency $f$, that the maximum of the mother wavelet $W(a,b)$ lies 
at $a=f_0/f$, so that the focii of the time-frequency
response of the signal to the wavelet transform are centered at frequencies 
$f=f_0/a$. For discretized time series data, $f=F_{\mathrm{s}} f_c / a$, where
$F_{\mathrm{s}}$ is the sample frequency of the data.  We determine 
empirically, via visual inspection, that a mother wavelet center
frequency $f_0=0.4$, and scales $a \in [1..128]$ yield a time-frequency 
resolution which allows us to resolve the pertinent substructure of the 
chirping features reported in this work.  We have used the {\tt pyCWT} software 
library available at~\cite{pycwt} to perform these decompositions numerically. \\

\subsection{Higher modes of the gravitational-wave emission:}

The complex \gw{} strain emitted in a direction $(\iota,\phi)$ on the sky of a BBH can be written as a superposition of different \gw{} modes $h_{\ell,m}(\iota,\phi;t)$ as \cite{Blanchet:2014zz}
\begin{equation}
h(t)=h_+(t) - \mathrm{i}h_{\times}(t)=\sum_{\ell \geq 2} \sum_{m = -\ell}^{m=\ell} Y_{\ell,m}^{-2}(\iota,\phi)h_{\ell,m}(t).
\end{equation} 
Here, $h_+$ and $h_\times$ denote the two \gw{} polarizations, the $Y_{\ell,m}$'s are spin$-2$ weighted spherical harmonics and $(\iota,\phi)$ are the polar and azimuthal angles of a spherical coordinate system centered on the binary.  This is chosen so that, $\iota=0$ (face-on) denotes the direction of the orbital angular momentum, while the orbital plane of the binary is located at $\iota=\mathrm{\pi}/2$ (edge-on). 
For face-on binaries, the quadrupolar $(\ell,m)=(2,\pm 2)$ modes vastly dominates during all the stages of the binary \cite{Blanchet:2014zz}. As consequence, the frequency of the resulting \gw{} is, to a good approximation, twice the orbital one and the observed signal shows a canonical \textit{single chirp} morphology. Secondary \gw{} modes are triggered during the merger and ringdown stages and are more visible near the orbital plane of the binary \cite{Berti:2009kk,Graff:2015bba,CalderonBustillo:2018zuq}. \\

\subsection{Computational methodology:}
We performed the simulations for this study using our \maya{} code~\cite{2007CQGHerrmann,2007PRDVaishnava,2009PRLHealy,2013PRDPekowsky}, a branch of the \etk{\cite{2012CQGLoffler}}. This code was also used produce the Georgia Tech catalog of gravitational waveforms~\cite{2016CQGJani}. We evolve the BSSN formulation of the Einstein equations~\cite{1998PRDBaumgarte} using the moving puncture gauge condition~\cite{2006PRLCampanelli,2006PRLBaker} for binary BH systems. The \maya{} code is built upon the \cactus{\cite{2003VPPR5ICGoodale}} code, using \kranc{\cite{2006CPCHusa}} for code generation and \carpet{\cite{2004CQGSchnetter}} for mesh refinement. We extract gravitational waveforms from the simulation data using the Newman-Penrose scalar $\Psi_4$~\cite{2010CQGReisswig}, which we calculate using the \textsc{WeylScal4} thorn of the \etk{.} We use the \ahf{\cite{2004CQGThornburg}} thorn with minor modifications (see below) to locate and analyze apparent horizons. \\

\subsection{Calculating curvatures on the apparent horizon:}
Consider an apparent horizon surface $\mathcal{S}$ (with 2-metric $\gamma_{ab}$) in a spacelike hypersurface $\Sigma_t$ (with 3-metric $g_{ij}$ and associated covariant derivative operator $\nabla_i$). The surface $\mathcal{S}$ can be defined by the outward-pointing unit normal to the surface $n^i$, whose divergence is equal to the mean curvature of the surface $H = \nabla_i n^i$. $H$ is calculated at run time by the \ahf{} thorn because it is an extrinsic quantity and therefore more difficult to calculate in post-processing, as it depends on the metric and its derivatives in the neighborhood of horizon. On the other hand, the Gaussian curvature $K$ of $\mathcal{S}$ is intrinsic to the surface and can therefore be calculated using only the metric induced on the horizon $\gamma_{ij} = g_{ij} - n_i n_j$ and its derivatives. We have modified the \ahf{} thorn to output $\gamma_{ij}$ on $\mathcal{S}$ for this purpose.

On a two-dimensional surface the Riemann tensor has only one independent component and is therefore completely determined by the Ricci scalar. The Gaussian curvature of $\mathcal{S}$ is half of the Ricci scalar and is related to components of the Riemann tensor by $R_{abcd} = K(\gamma_{ac}\gamma_{db} - \gamma_{ad}\gamma_{cb})$, where the indices $a$, $b$, $c$, and $d$ denote the polar ($\theta$) and azimuthal ($\phi$) angles on the horizon surface. Specifically, we use the $R_{\theta\phi\theta\phi}$ component of the Riemann tensor in our calculations. Furthermore, by calculating $K$ in the orbital plane of the binary we significantly reduce the complexity of this expression by taking advantage of the symmetry properties of $\gamma_{ab}$.\\

We compute derivatives with respect to the arc-length parameter $s$ on the intersection of the orbital plane and the apparent horizon. This is related to the azimuthal angle $\phi$ by

\begin{equation}
\frac{ds}{d\phi} = \sqrt{\gamma_{ij} \frac{dx^i}{d\phi}  \frac{dx^j}{d\phi}} = \sqrt{\gamma_{rr}\,(\partial_{\phi}R)^2 + \gamma_{\phi\phi}+2\gamma_{r\phi}\left(\partial_{\phi}R\right)}  \;.
\end{equation}

Here, $R$ denotes the coordinate radius of the horizon. The spherical components of $\gamma_{ij}$ are related to the cartesian components output by the \ahf{} thorn by:

\begin{equation}
\begin{aligned}
\gamma_{rr} & = \cos^2(\phi)\gamma_{xx} + \sin^2(\phi)\gamma_{yy} + \sin(2\phi)\gamma_{xy} \\
\gamma_{\phi\phi} & = r^2 \left[ \sin^2(\phi)\gamma_{xx} + \cos^2(\phi)\gamma_{yy} - \sin(2\phi)\gamma_{xy} \right]\\
\gamma_{r\phi} & = \left[\frac{1}{2} \sin(2\phi)(\gamma_{xx}-\gamma_{yy})+\cos(2\phi)\gamma_{xy}\right] 
\end{aligned}
\end{equation}\\

\end{methods}

\section{Supplementary Notes}

\subsection{Supplementary Note I. Mean curvature and Newman-Penrose scalar as a function of time:}

Similar to Fig. 5 in the main text, we show in Supplementary Fig. \ref{fig:curvature_and_dH_time} the values of the Newman-Penrose scalar $\Psi_4$ and the gradient of the mean curvature measured $dH/ds$ measured on the equator of the final apparent horizon of our $q=3$ binary, for different stages in its evolution. A similar correlation to that observed in Fig. 5 can be noted between $\Psi_4$ and $-dH/ds$. While we show here this correlation, in the main text we chose to show that between the absolute values of these quantities to facilitate a visual comparison to the corresponding simulation snapshots. \\

\subsection{Supplementary Note II. Varying mass ratio and gauge:}

Supplementary Figure.~\ref{fig:curvature_and_psi_all} demonstrates that the correlation presented in Figs. 4 and 5 in the main text for $q=3$ also hold for cases with different mass ratios. In particular we show cases with $q=2,5,6$ and $10$. As we acknowledge in the main text, the location of the apparent horizon itself depends of the choice of gauge. For this reason, we show results for the $q=2$ case obtained using two different coordinate gauge conditions: the standard moving puncture gauge condition~\cite{2006PRLCampanelli,2006PRLBaker} and a gauge with a dynamical shift condition~\cite{2010PRDMuller} that is more suitable for larger mass ratios. The moving puncture gauge uses a Gamma-driving shift condition with a damping parameter $\eta$ that has units of inverse mass. The range of appropriate values for $\eta$ depends on the mass of the \bh{s}, and thus for large mass ratios there is no one value that would lead to stability near both \bh{s}. In this case we use position-dependent value of $\eta$ to define a dynamical shift condition that is stable for larger mass ratios.\\

\subsection{Supplementary Note III. Observability of secondary chirps:}
Supplementary Figure~\ref{fig:distance} shows the distance at which the second chirp alone would produce a signal-to-noise ratio (SNR)  $\rho=5$. We assume that this secondary chirp is part of a longer confirmed observation with larger SNR, so that the noise may be assumed to be Gaussian \cite{Berry2015,LIGO_Guide}, making secondary chirp a $5\sigma$ deviation from the noise \cite{MatchedFilter,Abbott2020}. We consider four families of BBHs with varying mass ratio and total mass with the observer sitting on its orbital plane, $55^\circ$ away from the final kick direction. We assume two Advanced \ligo{} detectors working at both its current (solid) and design sensitivities (dashed) \cite{advLIGOcurves}. 
At design sensitivity, we find that a correctly oriented copy of GW170729, consistent with a mass ratio $q=2$ \cite{GWTC1,Chatziioannou2019}, would show an observable second chirp. We have checked that heavier asymmetric sources with $M = 300M_\odot$ would show visible secondary chirps up to distances of $\sim 6Gpc$.\\

\subsection{Calculation of sensitive distances:}

The optimal signal-to-noise ratio of a given signal $h(t)$ is defined as $\rho_{\text{opt}} = \sqrt{(h(t)|h(t))}$, where $(a|b)$ denotes the inner product defined as  \cite{Cutler1994}
\begin{equation}
(a|b)=4 \Re \int_{f_{min}}^{f_{cut}} \frac{\tilde{a}(f)\tilde{b}^{*}(f)}{S_n(f)} df.
\end{equation}
Here, $\tilde{a}(f)$ denotes the Fourier transform of $a(t)$, the asterisk denotes complex conjugation, $\Re$ denotes the real part and $S_n(f)$ represents the one-sided power spectral density of the detector noise. To compute the $\rho_{\text{opt}}$ of the second chirp alone, we cut our waveforms in the time domain, at the point where the time-frequency maps show the post-merger frequency minimum. To avoid Gibbs phenomena arising in the Fourier transforms of abruptly starting signals, we apply an aggressive window at the start of the already cut signal of width $\sim 10M$, so that our estimates of $\rho_{\text{opt}}$ are fairly conservative. We use a total mass-dependent lower frequency cutoff of $M \times f_0 = 0.015$, well below the lowest frequency for which the second chirp has support (see Fig.1 in the main text). The sensitive distance is then computed as the distance $D$ at which $h(t,D)$ produces $\rho_{\text{opt}} = 5$. We note that although this criterion guarantees that the observed signal would be a $5\sigma$ outlier from Gaussian noise, real \ligo{} noise is not Gaussian and an SNR of at least $\sim 10$ \textit{for the full signal} is usually needed to claim a gravitational-wave detection \cite{GWTC1}. We  assume that the secondary chirp is part of a longer signal that has been previously detected, so that noise may be safely assumed to be Gaussian \cite{Berry2015,LIGO_Guide}.  \\

\section{Supplementary Figures}

\begin{figure*}[hb!]
\begin{center}
\includegraphics[width=0.99\textwidth]{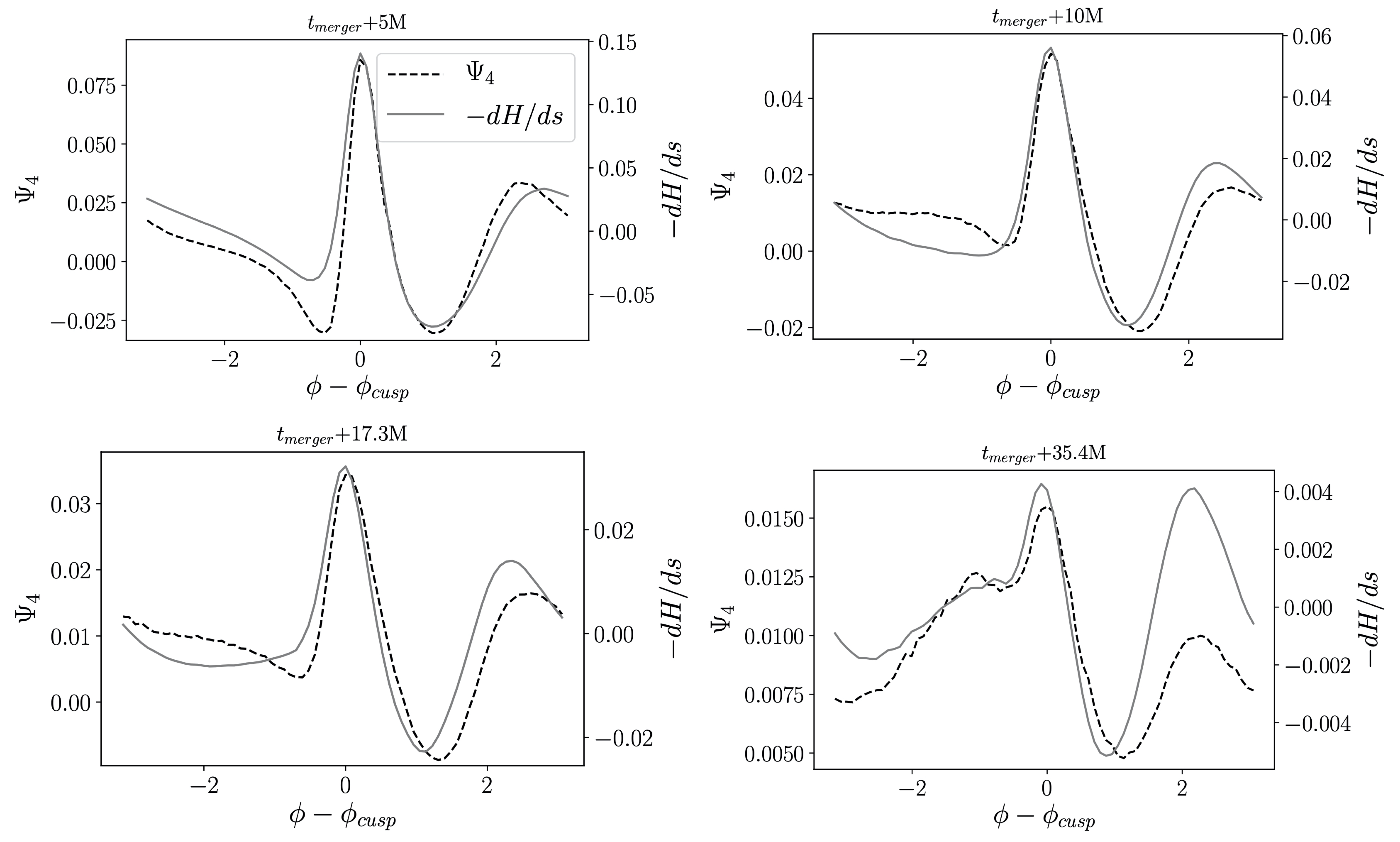}
\caption{\textbf{Relation between gravitational-wave emission and mean curvature on the apparent horizon}. Absolute values of the Newman-Penrose scalar $\Psi_4$ and the gradient of the mean curvature $dH/ds$ as a function of $\phi-\phi_{cusp}$ at times $5M$, $10M$, $17.3M$ and $35.4M$ after the formation of the common apparent horizon.}
\label{fig:curvature_and_dH_time}
\end{center}
\end{figure*}

\begin{figure*}
\begin{center}
\includegraphics[width=0.40\textwidth]{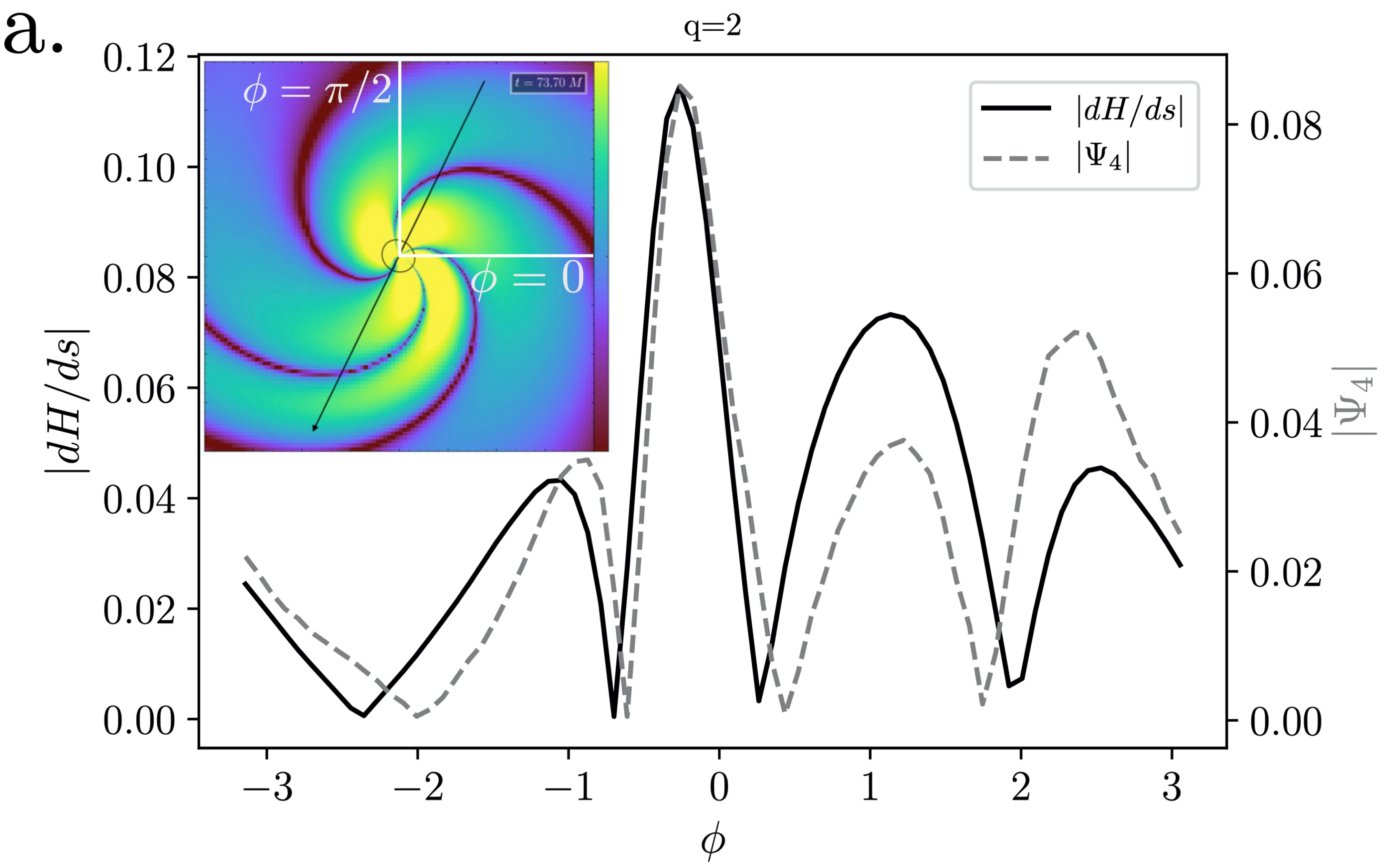}
\includegraphics[width=0.40\textwidth]{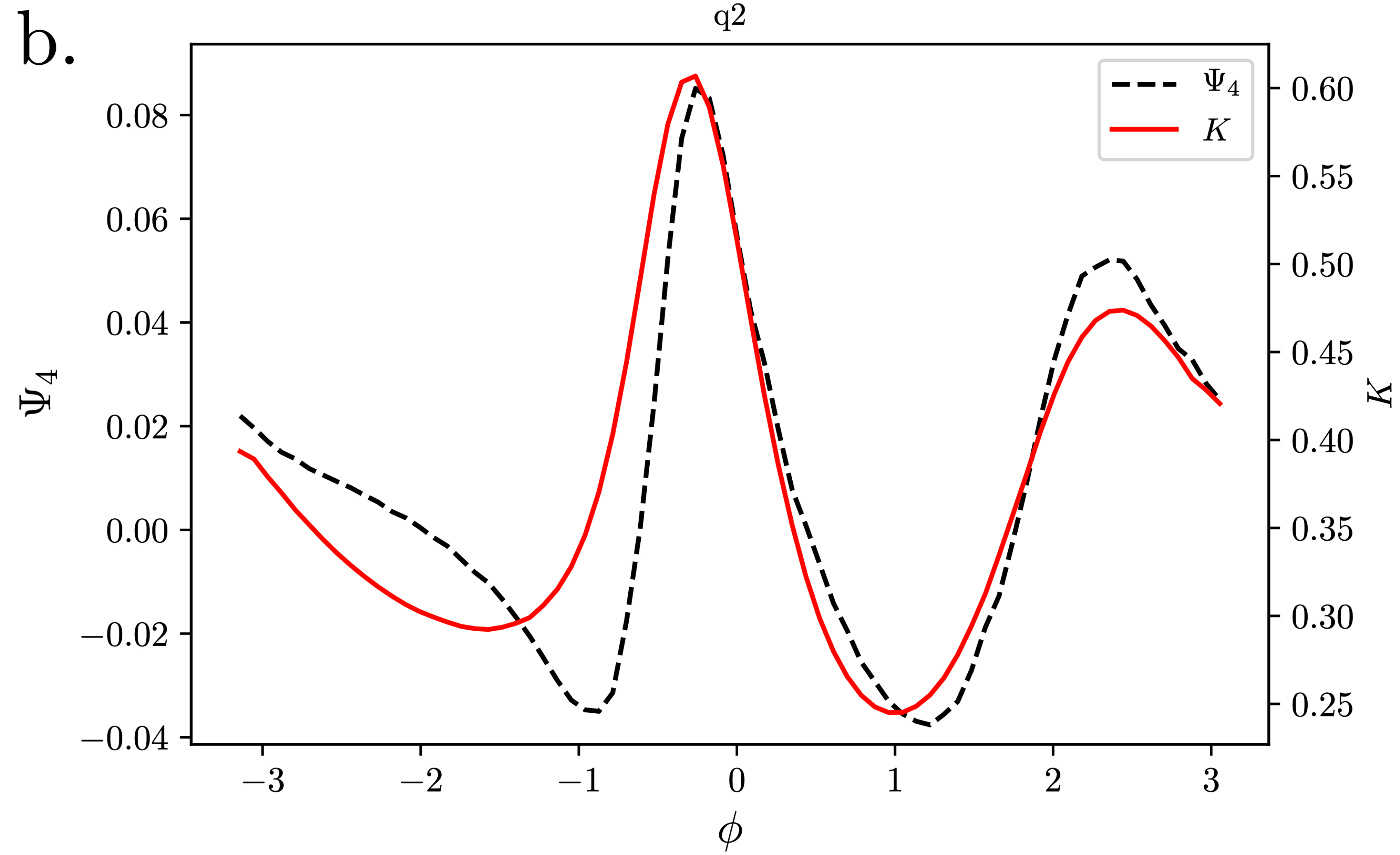}
\includegraphics[width=0.40\textwidth]{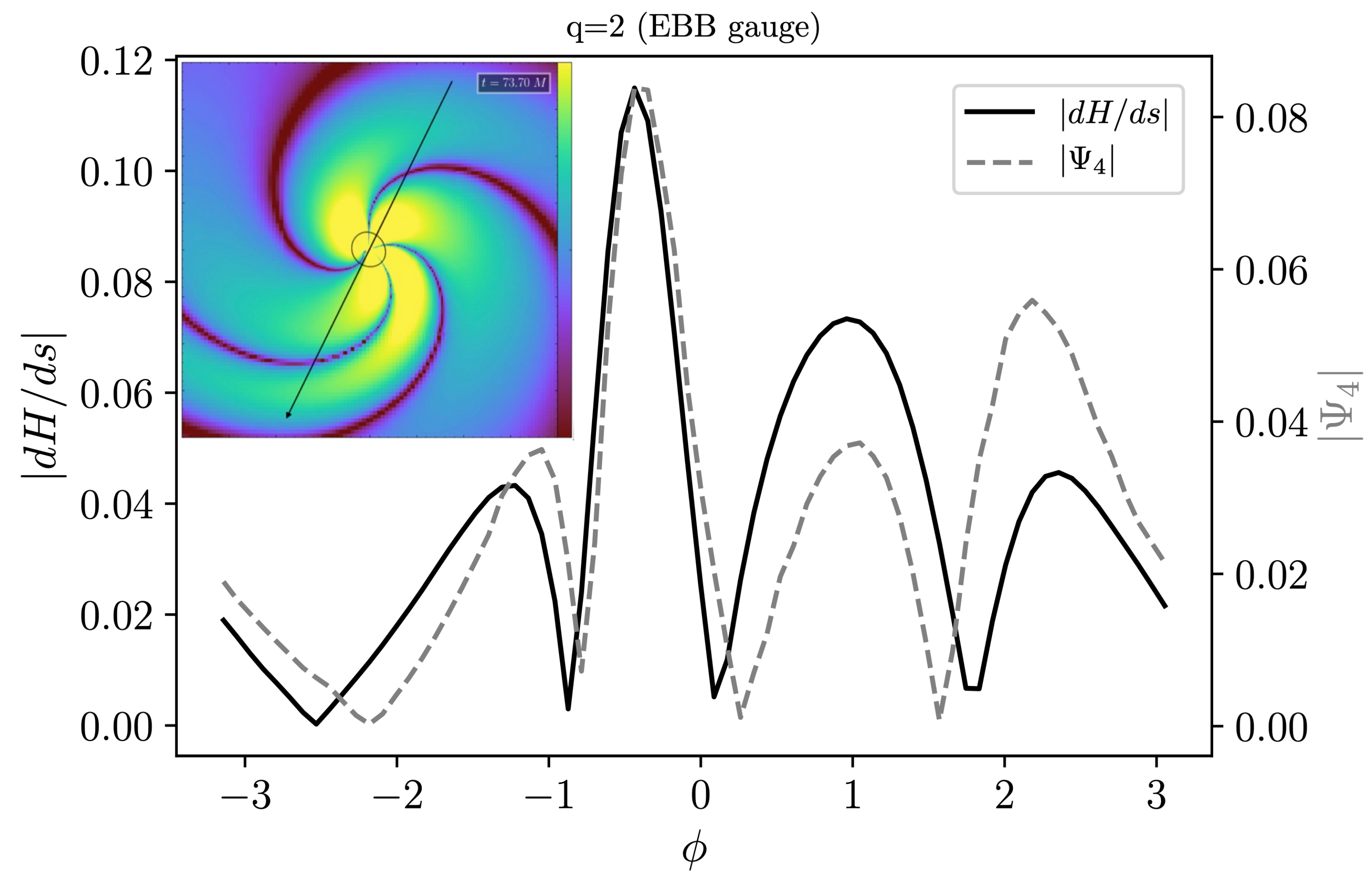}
\includegraphics[width=0.40\textwidth]{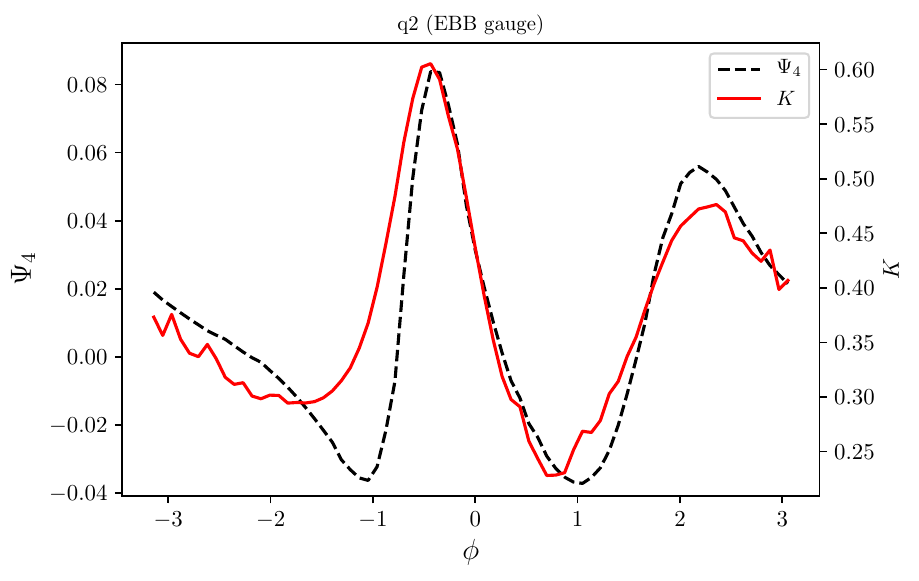}
\includegraphics[width=0.40\textwidth]{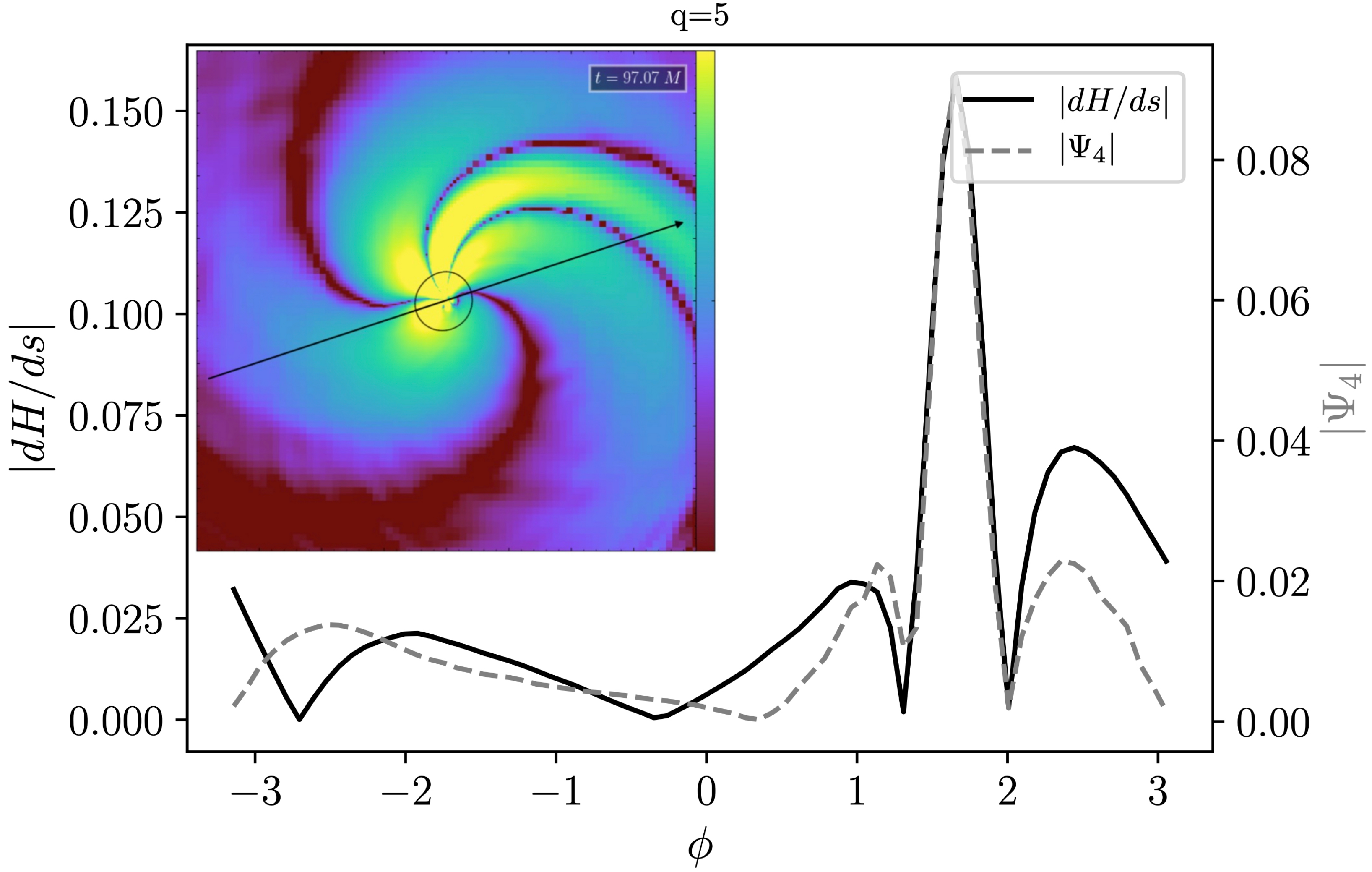}
\includegraphics[width=0.40\textwidth]{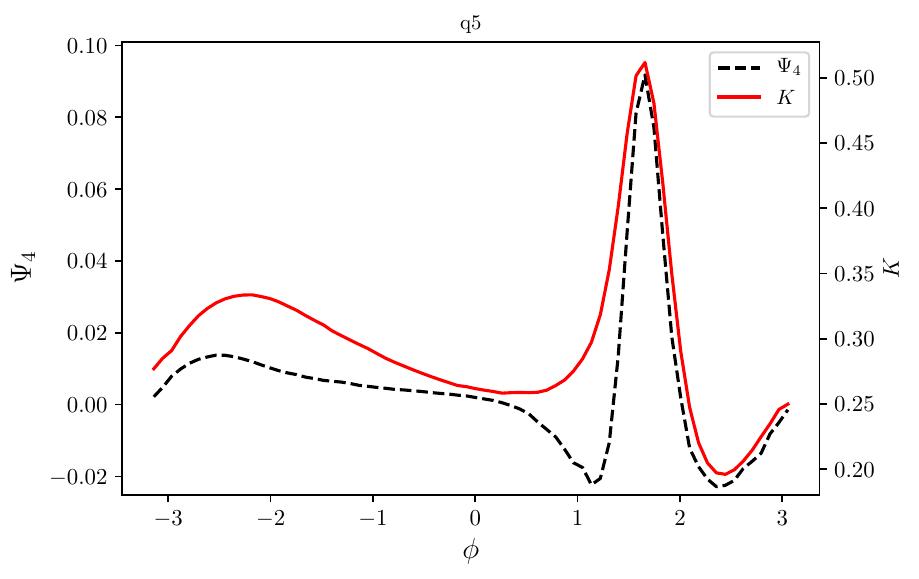}
\includegraphics[width=0.40\textwidth]{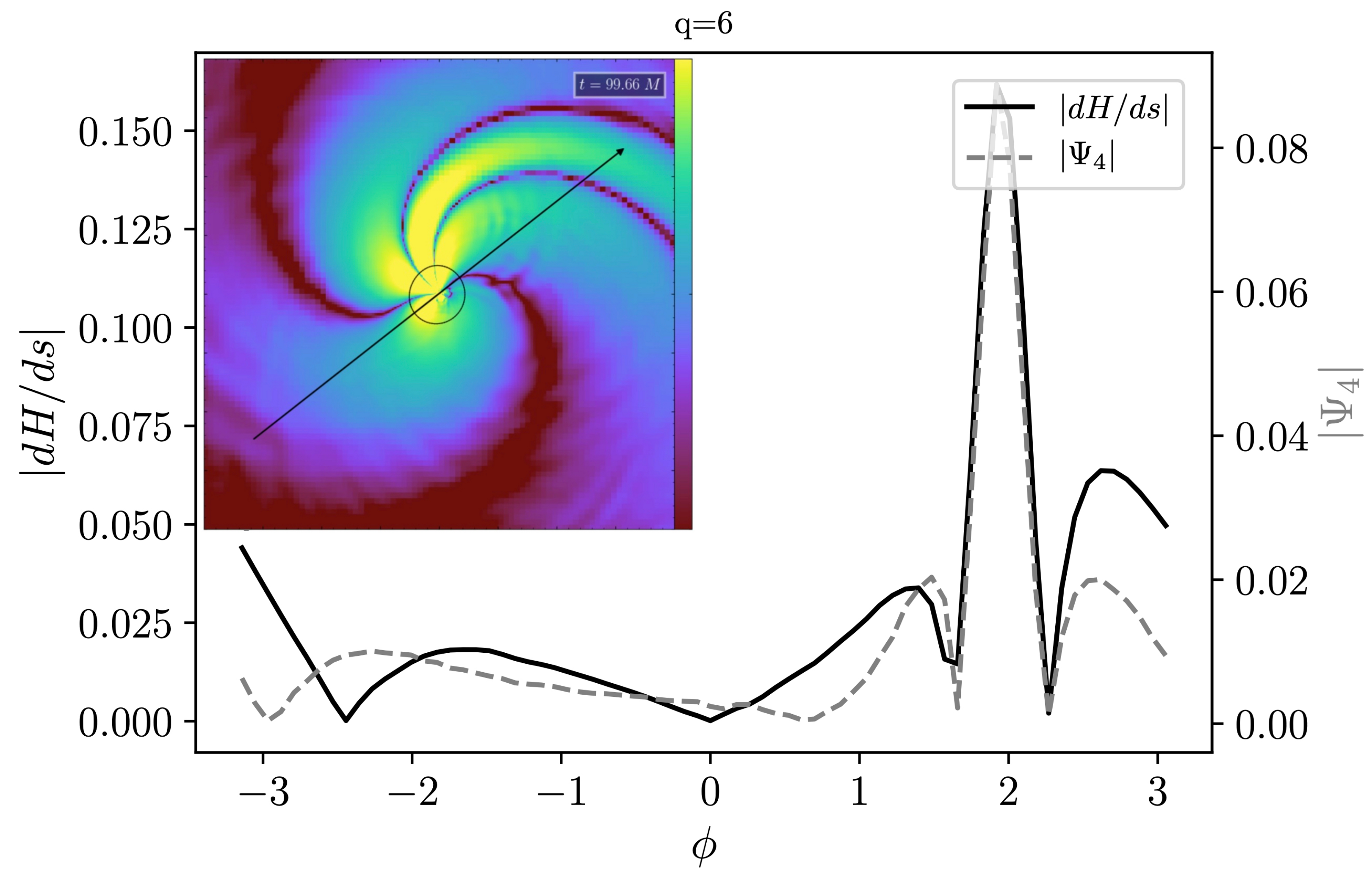}
\includegraphics[width=0.40\textwidth]{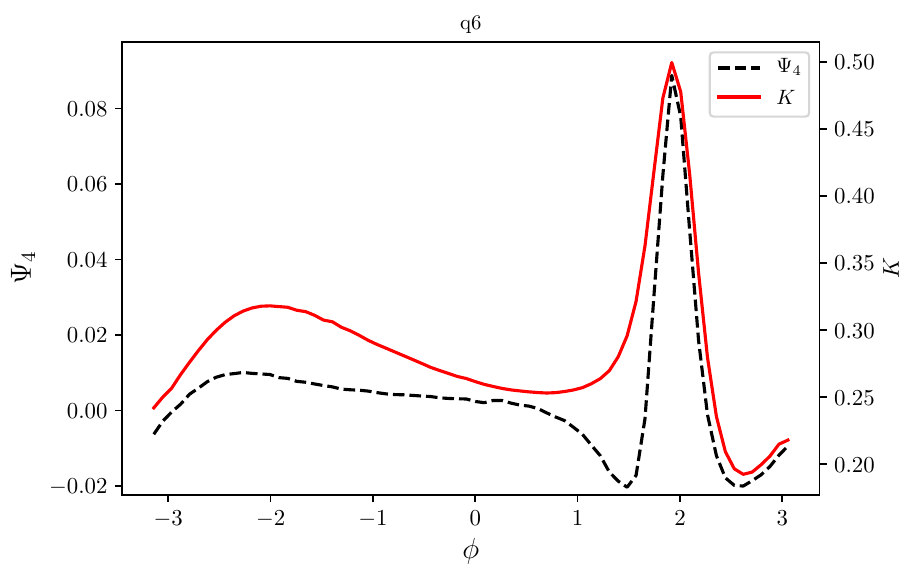}
\includegraphics[width=0.40\textwidth]{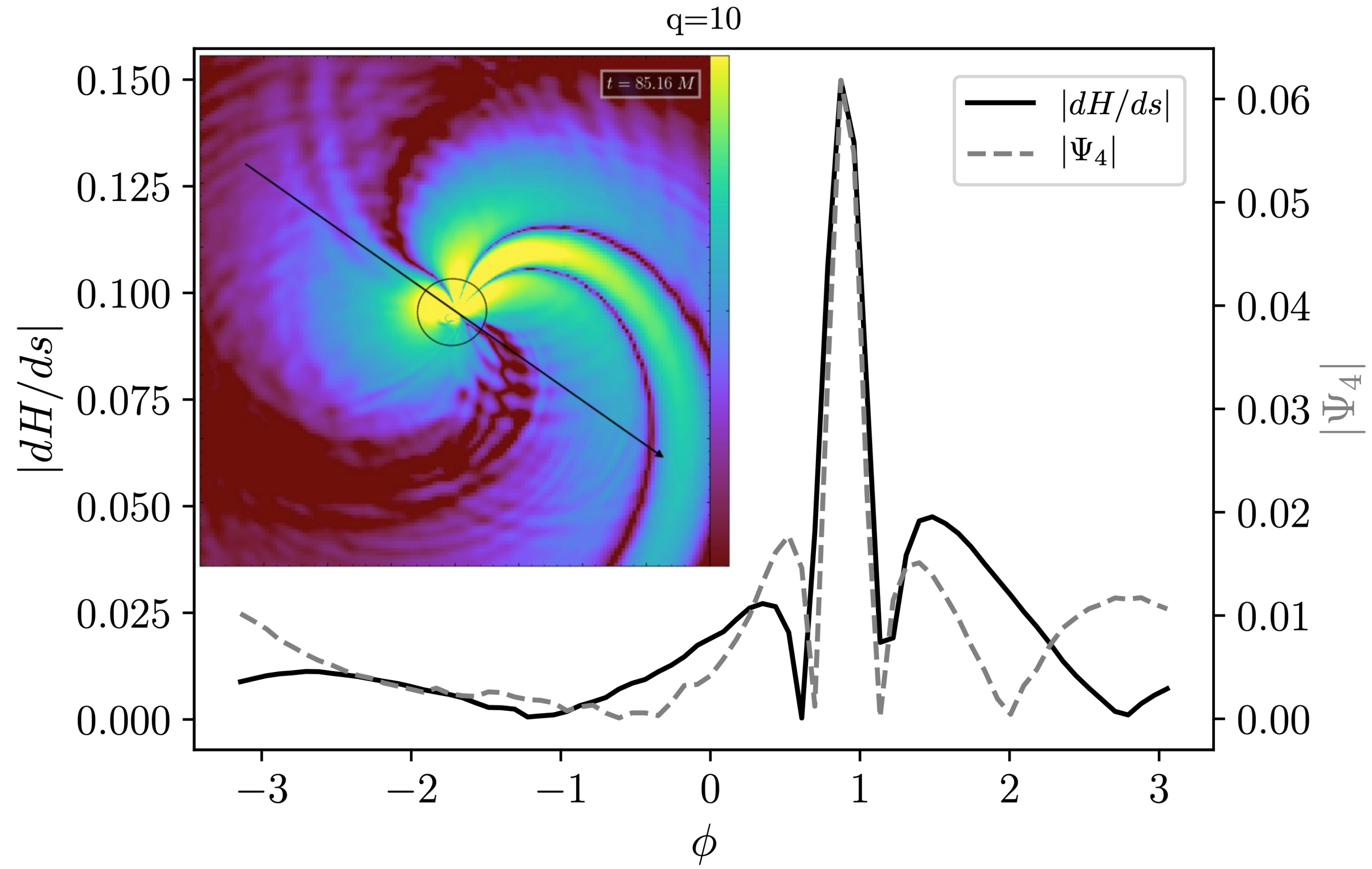}
\includegraphics[width=0.40\textwidth]{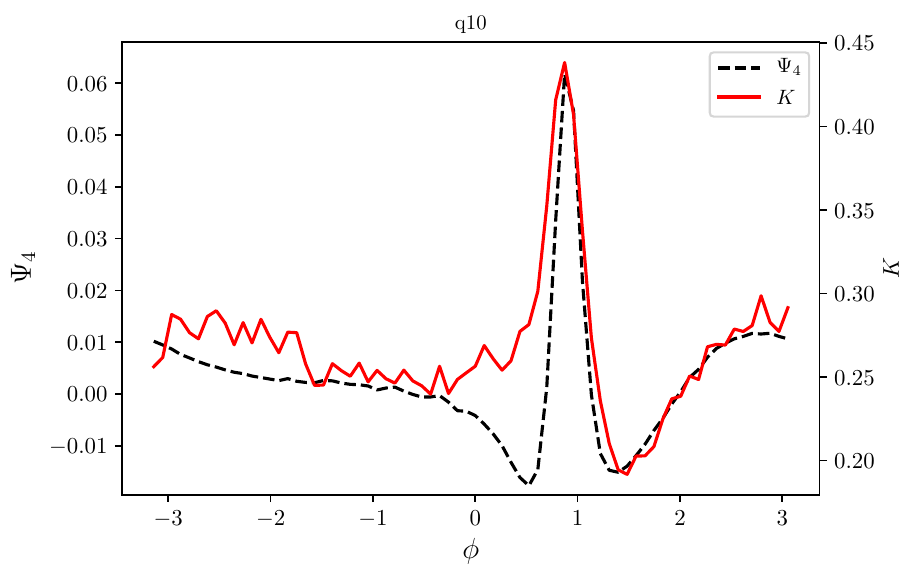}
\caption{\textbf{Relation between gravitational-wave emission and curvature for different mass ratio cases and different gauges}. Panels a show the value of the Newmann-Penrose scalar $\Psi_4$ and the gradient of the mean curvature $dH/ds$ along the intersection of the final apparent horizon and the original orbital plane of the binary for mass ratios $q=2,5,6$ and $10$. The two first panels show results for the $q=2$ case for two different coordinate gauge conditions. The inset shows $|\Psi_4|$ near the horizon. Panels b show the corresponding value of the Newman-Penrose scalar $\Psi_4$ and the Gaussian curvature $K$ for the same cases as panels a.}
\label{fig:curvature_and_psi_all}
\end{center}
\end{figure*}

\clearpage

\begin{center}
\begin{figure}[ht!]
\includegraphics[width=0.5\textwidth]{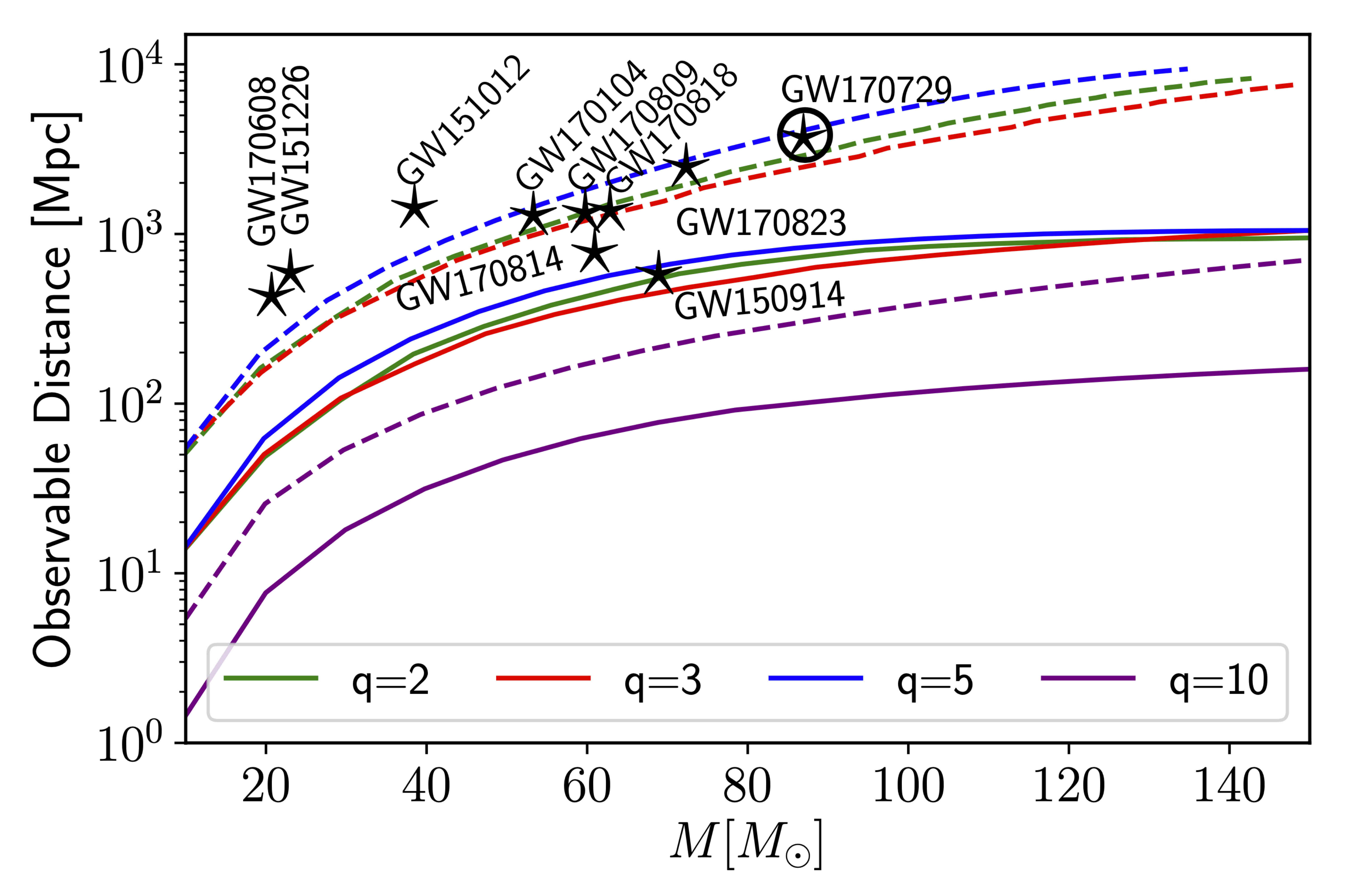}
\caption{\textbf{Observability of post-merger chirps}. Distance at which the secondary chirp from the remnant of several non-spinning binaries can produce a signal-to-noise ratio $\rho = 5$, as a function of the total mass of the binary measured in the detector frame. We consider two \ligo{} detectors working at its early sensitivity (solid) and design sensitivity (dashed). We consider only the case in which the secondary chirp is most intense, namely $(\iota,\phi) = (\pi/2, 11\pi/36)$ and choose optimal sky-location.}
\label{fig:distance}
\end{figure}
\end{center}

\section*{Data Availability} The Numerical Relativity waveforms in this work are part of the Georgia Tech catalogue \cite{2016CQGra..33t4001J} and are publicly available at http://www.einstein.gatech.edu/table. The waveforms used are, ordered by mass-ratio, are GT0717, GT0446, GT0453 ,GT0577, GT0604 and GT0568. The curvature and $\Psi_4$ data for the figures in this work are publicly available at https://github.com/cevans216/post-merger-chirp-data.

\section*{Code Availability} The numerical evolution code \maya{} is available upon request. The PyCBC and pyCWT codes are publicly available.

\section*{Acknowledgements}
We thank Paul D. Lasky and Valentin Christiaens for comments on the manuscript.
Work supported by NSF grants 1505824, 1505524, 1550461, XSEDE  TG-PHY120016. JCB also acknowledges support from Australian Research Council Discovery Project DP180103155 and the Direct Grant from the CUHK Research Committee with Project ID: 4053406. GK acknowledges support from the President's Undergraduate Research Salary Award of the Georgia Institute of Technology. Computational support through PACE at the Georgia Institute of Technology~\cite{PACE}. The results of Fig. 3 in our Supplementary material made use of the Python Numerical Relativity Injection infrastructure described in \cite{Schmidt:2017btt}. This manuscript has \ligo{} DCC number P1900139.

\section*{Author contributions} J.C.B. lead the project, noticed the multiple post-merger frequency peaks and its connection to the horizon curvature. C.E. performed the numerical simulations and developed the methods to compute quantities on the apparent horizon. J.C. provided the code to obtain time-frequency maps. G.K. developed the code used to compute sensitive distances. P.L. and D.S. provided key theoretical insights and supervised the work. All authors contributed to the writing and review of the manuscript.

\section*{Competing interests}  The Authors declare no Competing Financial or Non-Financial Interests
\section{Correspondence} $^{\dagger}$ juan.calderon.bustillo@gmail.com,  $^{\ddagger}$ cevans216@gatech.edu

\section*{References}
\bibliographystyle{naturemag}
\bibliography{IMBBH.bib}





\end{document}